\documentclass[epj]{svjour}

\usepackage{graphicx}
\usepackage{fancyhdr}
\usepackage{amsmath,amssymb}
\def\makeheadbox{{%  remove EPJC header
 }}
\def\mytitle{Dark Matter Candidates} 
\def\myauthors{Frank~Daniel~Steffen}  
\def\mytype{Review}
\def\mysession{Frank~Daniel~Steffen}
\newcommand{\Order}{{\cal O}}   % e.g. terms up to $\Order(g^2)$
\newcommand{\eV}{\mathrm{eV}}
\newcommand{\keV}{\mathrm{keV}}
\newcommand{\MeV}{\mathrm{MeV}}
\newcommand{\GeV}{\mathrm{GeV}}
\newcommand{\TeV}{\mathrm{TeV}}
\newcommand{\Mpc}{\mathrm{Mpc}}
\newcommand{\km}{\mathrm{km}}
\newcommand{\seconds}{\mathrm{s}}

\newcommand{\MPl}{\mathrm{M}_{\mathrm{P}}}
\newcommand{\proton}{\mathrm{p}}
\newcommand{\antiproton}{\bar{\mathrm{p}}}
\newcommand{\MET}{E_{T}^{\mathrm{miss}}}
\newcommand{\gravitino}{{\widetilde{G}}}
\newcommand{\axino}{{\widetilde{a}}}
\newcommand{\ax}{\ensuremath{\widetilde{a}}}
\newcommand{\stau}{{\widetilde{\tau}_1}}
\newcommand{\sel}{{\widetilde{e}_1}}
\newcommand{\smu}{{\widetilde{\mu}_1}}

\newcommand{\st}{{\tilde{\tau}_1}}
\newcommand{\bino}{{\widetilde B}}
\newcommand{\Bino}{{\widetilde B}}
\newcommand{\Wino}{{\widetilde W}}
\newcommand{\wino}{{\widetilde W}}

\newcommand{\neutralino}{{\widetilde \chi}^{0}_{1}}
\newcommand{\chargino}{{\widetilde{\chi}_1^{\pm}}}
\newcommand{\HiggsinoUp}{{\widetilde H}^{0}_{u}}
\newcommand{\HiggsinoDown}{{\widetilde H}^{0}_{d}}
\newcommand{\Higgs}{\mathrm{H}}
\newcommand{\mgr}{m_{\widetilde{G}}}

\newcommand{\mgravitino}{\mgr}
\newcommand{\mst}{m_{\tilde{\tau}_1}}
\newcommand{\mneu}{m_{{\widetilde \chi}^{0}_{1}}}

\newcommand{\LSP}{\mathrm{LSP}}
\newcommand{\NLSP}{\mathrm{NLSP}}
\newcommand{\LOSP}{\mathrm{LOSP}}
\newcommand{\NTP}{\mathrm{NTP}}
\newcommand{\TP}{\mathrm{TP}}

\newcommand{\equil}{\mathrm{eq}}

\newcommand{\freezeout}{\mathrm{f}}
\newcommand{\CDM}{\mathrm{dm}}

\newcommand{\EM}{\mathrm{em}}

\newcommand{\dec}{\mathrm{dec}}
\newcommand{\GUT}{\mathrm{GUT}}
\newcommand{\Reheating}{\mathrm{R}}
\newcommand{\TR}{T_{\Reheating}}

\newcommand{\Color}{\mathrm{c}}
\newcommand{\Weak}{\mathrm{L}}
\newcommand{\Hypercharge}{\mathrm{Y}}
\newcommand{\Lisix}{{}^6 \mathrm{Li}}
\newcommand{\Hefour}{{}^4 \mathrm{He}}

\newcommand{\taustau}{\tau_{\widetilde{\tau}_1}}

\newcommand{\monetwo}{m_{1/2}}
\newcommand{\mzero}{m_{0}}
\newcommand{\tanb}{\tan{\beta}}
\newcommand{\mgut}{M_\mathrm{GUT}}
\newcommand{\Omegatp}{\Omega_{\widetilde{G}}^{\mathrm{TP}}}
\newcommand{\Omegantp}{\Omega_{\widetilde{G}}^{\mathrm{NTP}}}

\newcommand{\champ}{X^{\! -}}

\newcommand{\be}{\begin{equation}}
\newcommand{\ee}{\end{equation}}
\newcommand{\bea}{\begin{eqnarray}}
\newcommand{\eea}{\end{eqnarray}}
\newcommand{\benn}{\begin{displaymath}}
\newcommand{\eenn}{\end{displaymath}}
\newcommand{\beann}{\begin{eqnarray*}}
\newcommand{\eeann}{\end{eqnarray*}}
\def\mytitle{Dark Matter Candidates} %Put your title here!
\def\myauthors{Frank~Daniel~Steffen}    %Put your name here!
\def\mytype{Review}
\def\mysession{\myauthors}
\begin{document}
\title{Supersymmetric Dark Matter Candidates}
\subtitle{The lightest neutralino, the gravitino, and the axino}
\author{Frank Daniel Steffen\inst{1}
\thanks{\emph{Email:} steffen@mppmu.mpg.de}%
}                     % Do not remove
\institute{Max-Planck-Institut f\"ur Physik,
F\"ohringer Ring 6,
D-80805 Munich,
Germany}
\date{}
\abstract{
  In supersymmetric extensions of the Standard Model, the lightest
  neutralino, the gravitino, and the axino can appear as the lightest
  supersymmetric particle and as such provide a compelling explanation
  of the non-baryonic dark matter in our Universe.
  For each of these dark matter candidates, I review the present
  status of primordial production mechanisms, cosmological
  constraints, and prospects of experimental identification.
 % end of PACS codes
} %end of abstract
%
%
% 04.65.+e 	Supergravity (see also 12.60.Jv Supersymmetric models)
%% 12.60.Jv 	Supersymmetric models (see also 04.65.+e Supergravity)
% 14.80.Ly 	Supersymmetric partners of known particles
%% 95.30.Cq 	Elementary particle processes
%% 95.35.+d 	Dark matter
%% 98.80.Cq 	Particle-theory and field-theory models of the early Universe
%
\maketitle
%DO NOT REMOVE THIS LINE

% __________________________________________________________________
\section{Introduction}
\label{intro}
% __________________________________________________________________

Numerous astrophysical and cosmological considerations point to the
existence of non-baryonic dark matter in our
Universe~\cite{Bergstrom:2000pn,Bertone:2004pz}.
In fact, based on observations of supernovae, galaxy clusters, and the
cosmic microwave background (CMB), we believe today that our Universe
is flat with about 76\%, 20\%, and 4\% of the critical energy density
$\rho_c$ provided in the form of dark energy, non-baryonic dark
matter, and baryons, respectively~\cite{Spergel:2006hy,Yao:2006px}.
A nominal ``$3\sigma$'' range%
\footnote{Note that the nominal ``$3\sigma$'' range is derived
  assuming a restrictive six-parameter ``vanilla'' model. A larger
  range is possible---even with additional data from other
  cosmological probes---if the fit is performed in the context of a
  more general model that includes other physically motivated
  parameters such as a nonzero neutrino mass~\cite{Hamann:2006pf}.
  Thereby, the range $0.094 < \Omega_{\CDM} h^2 < 0.136$ has been
  obtained in Ref.~\cite{Hamann:2006pf}.}
of the dark matter density $\Omega_{\CDM}=\rho_{\CDM}/\rho_c$ can be
inferred from measurements of the CMB anisotropies by the Wilkinson
Micro\-wave An\-iso\-tropy Probe (WMAP)
satellite~\cite{Spergel:2006hy}
\begin{equation}
        \Omega_{\CDM}^{3\sigma}h^2=0.105^{+0.021}_{-0.030} 
%        \ ,
\label{Eq:OmegaDM}
\end{equation} 
with $h=0.73^{+0.04}_{-0.03}$ denoting the Hubble constant in units of
$100\,\km\,\Mpc^{-1}\seconds^{-1}$.

Relying on the pieces of evidence, we think that a particle physics
candidate for dark matter has to be electrically neutral, color
neutral,%
\footnote{A colored dark matter candidate is disfavored by severe
  limits from searches for anomalous heavy nuclei~\cite{Yao:2006px}.}
and stable or have a lifetime $\tau_{\CDM}$ that is not much smaller
than the age of the Universe today $t_0 \simeq 14~\mathrm{Gyr}$.
Moreover, the species providing the dominant contribution to
$\Omega_{\CDM}$ have to be sufficiently slow to allow for structure
formation. For example, since the neutrinos of the Standard Model are
too light, $\sum_i m_{\nu_i} \lesssim
\Order(1~\eV)$~\cite{Lesgourgues:2006nd}, they were too fast at early
times. Accordingly, they are classified as hot dark matter which can
constitute only a minor fraction of $\Omega_{\CDM}$ since otherwise
structure formation cannot be understood~\cite{Hannestad:2007dd}.
Thus, the observationally inferred dark matter density can be
considered as evidence for physics beyond the Standard Model.

Supersymmetric (SUSY) extensions of the Standard Model are an
appealing concept because of their remarkable properties, for example,
with respect to gauge coupling unification, the hierarchy problem, and
the embedding of
gravity~\cite{Wess:1992cp,Nilles:1983ge,Haber:1984rc,Martin:1997ns,Drees:2004jm,Baer:2006rs}.
As superpartners of the Standard Model particles, new particles appear
including fields that are electrically neutral and color neutral.
Since they have not been detected at particle accelerators, these
sparticles must be heavy or extremely weakly interacting.

Because of the non-observation of reactions that violate lepton number
$L$ or baryon number $B$, it is often assumed---as also in this
review---that SUSY theories respect the multiplicative quantum number 
\begin{align}
  \mathrm{R}=(-1)^{3B+L+2S} 
  \ ,
\end{align}
known as R-parity, with $S$ denoting the spin. Since Standard Model
particles and superpartners carry respectively even (+1) and odd (-1)
R-parity, its conservation implies that superpartners can only be
produced or annihilated in pairs and that the lightest supersymmetric
particle (LSP) cannot decay even if it is heavier than most (or all)
of the Standard Model particles.%
\footnote{While R-parity conservation is assumed in this review, its
  violation is a realistic option; see, e.g.,
  \cite{Dreiner:1997uz,Allanach:2007vi,Takayama:2000uz,Buchmuller:2007ui,Ibarra:2007jz}.}
An electrically neutral and color neutral LSP can thus be a compelling
dark matter candidate. For the lightest neutralino, the gravitino, and
the axino, which are well-motivated LSP candidates, this is shown
below. For each scenario, I will address implications for cosmology
and experimental prospects.
Note that the discussion of gravitino/axino dark matter in
Sects.~\ref{sec:GravitinoDM} and~\ref{sec:AxinoDM} will be more
extensive than the one of neutralino dark matter in
Sect.~\ref{sec:NeutalinoDM}, for which numerous excellent reviews
exist such
as~\cite{Jungman:1995df,Drees:2004jm,Olive:2007hm,Bertone:2007ki}.

% __________________________________________________________________
\section{Neutralino Dark Matter}
\label{sec:NeutalinoDM}
% __________________________________________________________________

The lightest neutralino $\neutralino$ appears already in the minimal
supersymmetric Standard Model (MSSM) as the lightest mass eigenstate
among the four neutralinos being mixtures of the bino $\bino$, the
wino $\wino$, and the neutral higgsinos $\HiggsinoUp$ and
$\HiggsinoDown$. Accordingly, $\neutralino$ is a spin 1/2 fermion with
weak interactions only. Its mass $\mneu$ depends on the gaugino mass
parameters $M_{1}$ and $M_{2}$, on the ratio of the two MSSM Higgs
doublet vacuum expectation values $\tanb$, and the higgsino mass
parameter $\mu$. Expecting $\mneu=\Order(100~\GeV)$, $\neutralino$ is
classified as a weakly interacting massive particle (WIMP).

Motivated by theories of grand unification and
supergravity~\cite{Brignole:1997dp} and by experimental constraints on
flavor mixing and CP violation~\cite{Yao:2006px}, one often assumes
universal soft SUSY breaking parameters at the scale of grand
unification $\mgut$;
cf.~\cite{Martin:1997ns,Drees:2004jm,Baer:2006rs,Olive:2007hm} and
references therein.  For example, in the framework of the constrained
MSSM (CMSSM), the gaugino masses, the scalar masses, and the trilinear
scalar interactions are assumed to take on the respective universal
values $\monetwo$, $\mzero$, and $A_0$ at $\mgut$. Specifying
$\monetwo$, $\mzero$, $A_0$, $\tanb$, and the sign of $\mu$, the
low-energy mass spectrum is given by the renormalization group running
from $\mgut$ downwards.

Assuming $A_0=0$ for simplicity, the lightest Standard Model
superpartner---or lightest ordinary superpartner (LOSP)---is either
the lightest neutralino $\neutralino$ or the lighter stau $\stau$,
whose mass is denoted by $m_{\stau}$. If the LSP is assumed to be the
LOSP, the parameter region in which $m_{\stau}<m_{\neutralino}$ is
usually not considered because of severe upper limits on the abundance
of stable charged particles~\cite{Yao:2006px}. However, in
gravitino/axino LSP scenarios, in which the LOSP is the
next-to-lightest supersymmetric particle (NLSP), the $\stau$ LOSP case
is viable and particularly promising for collider phenomenology as
will be discussed in Sects.~\ref{sec:GravitinoDM}
and~\ref{sec:AxinoDM}.

\begin{figure}[t!]
\includegraphics[width=0.45\textwidth]{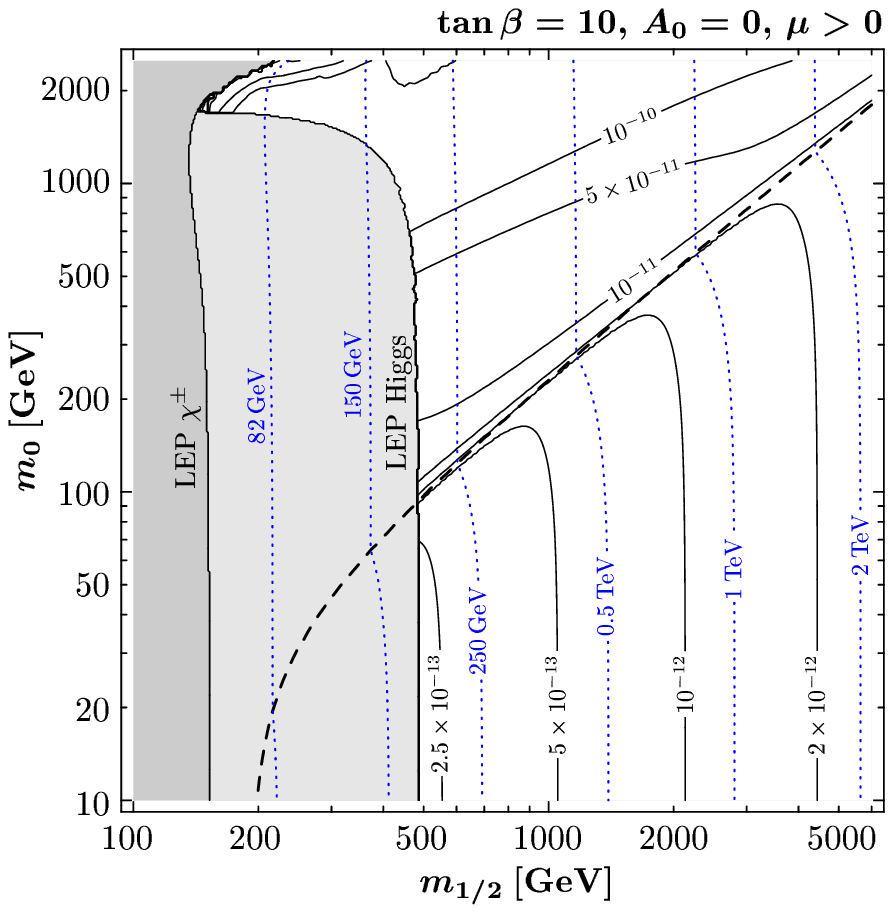} 
\caption{Contours of $m_{\LOSP}$ (dotted blue lines) and
  $Y_{\LOSP}^{\dec}$ (solid black lines) in the $(m_{1/2},m_0)$ plane
  for $A_0=0$, $\mu>0$, $\tan\beta=10$. Above (below) the dashed line,
  $m_{\neutralino}<m_{\stau}$ ($m_{\stau}<m_{\neutralino}$). The
  medium gray and the light gray regions show the LEP bounds
  $m_{\chargino}>94~\GeV$ and $m_{\Higgs}>114.4~\GeV$,
  respectively~\cite{Yao:2006px}. The contours are obtained with the
  spectrum generator {\tt SuSpect~2.34}~\cite{Djouadi:2002ze} using
  $m_t=172.5~\GeV$ and
  $m_{\mathrm{b}}(m_{\mathrm{b}})^{\mathrm{\overline{MS}}} = 4.23\
  \GeV$, and with {\tt
    micrOMEGAs~1.37}~\cite{Belanger:2001fz,Belanger:2004yn}.
  From~\cite{Pradler:2006hh}.}
\label{Fig:YLOSP}
\end{figure}

In Fig.~\ref{Fig:YLOSP} (from~\cite{Pradler:2006hh})
%
% __________________________________________________________________
% __________________________________________________________________
%
the dotted (blue in the web version) lines show contours of
$m_{\LOSP}$ in the $(m_{1/2},m_0)$ plane for $A_0=0$, $\mu>0$,
$\tan\beta=10$. Above (below) the dashed line,
$m_{\neutralino}<m_{\stau}$ ($m_{\stau}<m_{\neutralino}$). The medium
gray and the light gray regions at small $m_{1/2}$ are excluded
respectively by the mass bounds $m_{\chargino}>94~\GeV$ and
$m_{\Higgs}>114.4~\GeV$ from chargino and Higgs searches at
LEP~\cite{Yao:2006px}. It can be seen that $\mneu=\Order(100~\GeV)$
appears naturally within the CMSSM.

% __________________________________________________________________
\subsection{Primordial Origin}
\label{sec:NeutralinoProduction}
% __________________________________________________________________

The $\neutralino$'s were in thermal equilibrium for primordial
temperatures of $T>T_{\freezeout}\simeq\mneu/20$. At $T_{\freezeout}$,
the annihilation rate of the (by then) non-relativistic
$\neutralino$'s becomes smaller than the Hubble rate so that they
decouple from the thermal plasma.  Thus, for $T\lesssim
T_{\freezeout}$, their yield $Y_{\neutralino}\equiv n_{\neutralino}/s$
is given by
$Y_{\neutralino}^{\dec}\approx
Y^{\equil}_{\neutralino}(T_{\freezeout})$, 
where $n_{\neutralino}^{(\equil)}$ is the (equilibrium) number density
of $\neutralino$'s and $s = 2\pi^2\,g_{*S}\,T^3/45$ the entropy
density. Depending on details of the $\neutralino$ decoupling,
$Y_{\neutralino}^{\dec}$ is very sensitive to the mass spectrum and
the couplings of the superparticles.  Indeed, convenient computer
programs such as {\tt DarkSUSY}~\cite{Gondolo:2004sc} or {\tt
  micrOMEGAs 1.37}~\cite{Belanger:2001fz,Belanger:2004yn} are
available which allow for a numerical calculation of the LOSP
decoupling and the resulting thermal relic abundance in a given SUSY
model.

\begin{figure}[t!]
%\hspace*{.5cm} 
${m_0}$ [{\bf GeV}]
\begin{center}
\includegraphics[width=.45\textwidth,height=.4\textwidth]{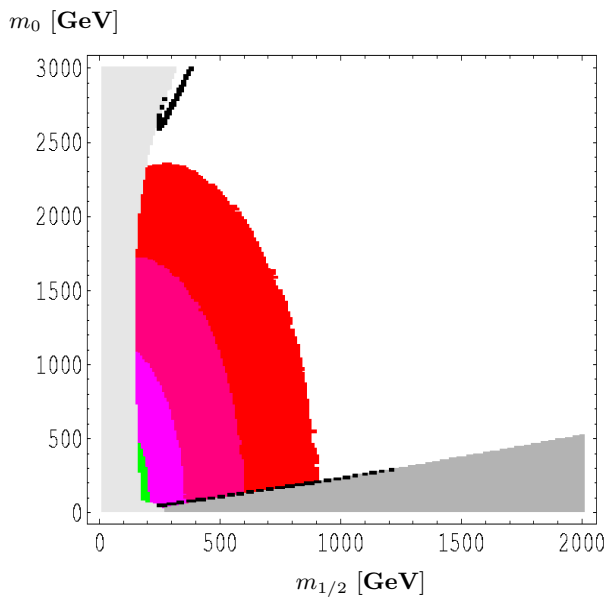}
\hspace*{1.cm} $m_{1/2}$ [{\bf GeV}]
\end{center}
\caption{Regions (black) with $0.087 \leq\Omega_{\neutralino}h^2\leq
  0.138$ in the $(m_{1/2},m_0)$ plane for $A_0=0, \mu>0$,
  $\tan\beta=10$, and $m_t=172.7~\GeV$. In the dark gray triangular
  region, $\mneu>m_{\stau}$. The light gray region at small $m_{1/2}$
  is excluded by the requirement of correct electroweak symmetry
  breaking or by sparticle search limits~\cite{Djouadi:2006be}, the
  two medium shaded (light pink in the web version) bands by the LEP
  bound $m_{\Higgs}>114~\GeV$, and the small light shaded (green in
  the web version) spot by the $b \rightarrow s \gamma$ constraint:
  $2.65\leq \mathrm{BR}(b \rightarrow s \gamma)/10^{-4} \leq 4.45$.
  The dark shaded (red in the web version) band is compatible with
  having a Standard Model like Higgs boson near 115~GeV.
  From~\cite{Djouadi:2006be}.}
\label{Fig:NeutralinoDM}
\end{figure}

The $Y_{\LOSP}^{\dec}$ contours shown by the solid black lines in
Fig.~\ref{Fig:YLOSP} illustrate that the $\neutralino$ LSP yield can
easily vary by more than an order of magnitude. Because of this
sensitivity, the associated thermal relic density
\begin{equation}
  \Omega_{\neutralino} h^2
  = \mneu\,Y_{\neutralino}^{\dec}\,s(T_0)\,h^2/\rho_c
\label{Eq:NeutralinoDensity}
\end{equation}
agrees with $\Omega_{\CDM}^{3\sigma}h^2$ only in narrow regions in the
parameter space; $\rho_c/[s(T_0)h^2]=3.6\times
10^{-9}\,\GeV$~\cite{Yao:2006px}.  This can be seen in
Fig.~\ref{Fig:NeutralinoDM} (from~\cite{Djouadi:2006be})
%
% __________________________________________________________________
% __________________________________________________________________
%
where the black strips indicate the region with $0.087 \leq
\Omega_{\neutralino}h^2\leq 0.138$.

Remarkably, it is exactly the small width of the
$\Omega_{\neutralino}=\Omega_{\CDM}$ regions which could help us to
identify $\neutralino$ dark matter. Once sparticles are produced at
colliders, the data analysis will aim at determinig the SUSY model
realized in nature~\cite{Lafaye:2004cn,Bechtle:2004pc}. For the
reconstructed model, a precise calculation of $\Omega_{\neutralino}$
is possible assuming a standard thermal history of the Universe.
Because of the sensitivity of $\Omega_{\neutralino}$ with respect to
the SUSY model, an agreement of the obtained $\Omega_{\neutralino}$
with $\Omega_{\CDM}$ will then be strong evidence for the
$\neutralino$ LSP providing $\Omega_{\CDM}$ and for a standard thermal
history up to the $\neutralino$-decoupling temperature $T_{\freezeout}$. Since
$\neutralino$'s decouple already as a non-re\-la\-ti\-vis\-tic
species, it is also guaranteed that they are sufficiently cold to
allow for cosmic structure formation.

% __________________________________________________________________
\subsection{Experimental Prospects}
\label{sec:NeutralinoExperiments}
% __________________________________________________________________

For experimental tests of the $\neutralino$ dark matter hypothesis,
three complementary techniques exist: indirect, direct, and collider
searches.  While there is an enormous activity in each of those
fields, I will summarize only the main ideas. For more detailed
discussions, see~\cite{Baltz:2006fm,Bertone:2007ki,Baudis:2007dq} and
references therein.

Let us first turn to indirect searches. Since dark matter clumps, one
expects regions with an increased $\neutralino$ density such as galaxy
halos, the center of galaxies, and the center of stars.  While
$\neutralino$ pair annihilation after $\neutralino$ decoupling is
basically negligible for calculations of $\Omega_{\neutralino}$, it
should occur at a significant rate in these regions.  The resulting
Standard Model particles should then lead to energetic cosmic rays and
thereby to an excess of photons, neutrinos, positrons, and antiprotons
over backgrounds expected from standard cosmic ray models without dark
matter annihilation.  In fact, data from the Energetic Gamma Ray
Experiment Telescope (EGRET) has already been interpreted as evidence
for $\neutralino$ annihilation~\cite{deBoer:2005bd,Elsaesser:2004ap}
within SUSY models that will be testable in direct and collider
searches.  For a discussion of these and other potential hints,
see~\cite{Hooper:2007vy,Bertone:2007ki} and references therein.

In direct searches, one looks for signals of $\neutralino$'s---or more
generally WIMPs---passing through earth that scatter elastically off
nuclei. Being located in environments deep underground that are well
shielded against unwanted background, an enormous sensitivity has
already been
reached~\cite{Angloher:2004tr,Sanglard:2005we,Akerib:2005kh,Angle:2007uj}.
Since no unambiguous signal of a $\neutralino$--nucleus scattering
event has been observed so far, $\mneu$-dependent upper limits on the
respective $\neutralino$ cross section are obtained. Indeed, the
current best limits given by the CDMS~II~\cite{Akerib:2005kh} and the
Xenon~10~\cite{Angle:2007uj} experiments exclude already a part of the
SUSY parameter space; see, for
example,~\cite{Olive:2007hm,Bertone:2007ki,Baudis:2007dq} and
references therein.  These limits, however, depend on the assumed
$\neutralino$ flux at the detector location, which is subject to
significant uncertainties due to possible inhomogeneities in the dark
matter distribution in galaxies. Such inhomogeneities should manifest
themselves also in indirect searches which can help to reduce those
uncertainties.  Once $\neutralino$ events are observed in direct
searches, one can succeed in reconstructing the $\neutralino$ velocity
distribution~\cite{Drees:2007hr}. By analyzing the recoil spectra,
$\mneu$ can even be estimated in a way that is independent of the dark
matter density on earth~\cite{Shan:2007vn}.

In most searches for SUSY at colliders, it is assumed that R-parity is
conserved. Accordingly, one expects that superpartners are produced in
pairs before decaying via cascades into the LSP and energetic
fermions. As a weakly-interacting particle, every $\neutralino$ LSP
produced will escape the detector without leaving a track. Thus, the
existence of SUSY and the $\neutralino$ LSP has to be inferred from
studies of missing transverse energy $\MET$ and energetic jets and
leptons emitted along the cascades. Along these lines, ongoing
investigations are pursued based on data from $\proton\antiproton$
collisions with a center-of-mass energy of $\sqrt{s}=2~\TeV$ at the
Fermilab Tevatron Collider. While lower limits on the masses of
squarks and gluinos have been extracted, no evidence for SUSY or the
$\neutralino$ LSP has been reported so
far~\cite{Duperrin:2007uy,Shamim:2007yy}.
With the first $\proton\proton$ collisions with $\sqrt{s}=14~\TeV$ at
the CERN Large Hadron Collider (LHC) expected in the year 2008, there
are high hopes that the new energy range will allow for a copious
production of superpartners. Here large $\MET$ will be the key
quantity for early SUSY searches~\cite{Tytgat:2007gj,Yamamoto:2007it}.
Despite an enormous potential for mass and spin measurements of SUSY
particles at the LHC~\cite{Ozturk:2007ap}, additional precision
studies at the planned International Linear Collider
(ILC)~\cite{Weiglein:2004hn,Choi:2007zg} appear to be crucial for the
identification of the $\neutralino$
LSP~\cite{Baltz:2006fm,Choi:2006mr}.

% __________________________________________________________________
\section{Gravitino Dark Matter}
\label{sec:GravitinoDM}
% __________________________________________________________________

The gravitino $\gravitino$ appears (as the spin-3/2 superpartner of
the graviton) once SUSY is promoted from a global to a local symmetry
leading to supergravity~\cite{Wess:1992cp}. The gravitino
mass~$m_{\gravitino}$ depends strongly on the SUSY-breaking scheme and
can range from the eV scale to scales beyond the TeV
region~\cite{Nilles:1983ge,Martin:1997ns,Dine:1994vc,Dine:1995ag,Giudice:1998bp,Randall:1998uk,Giudice:1998xp,Buchmuller:2005rt}.
For example, in gauge-mediated SUSY breaking
schemes~\cite{Dine:1994vc,Dine:1995ag,Giudice:1998bp}, the mass of the
gravitino is typically less than 1~GeV, while in gravity-mediated
sche\-mes~\cite{Nilles:1983ge,Martin:1997ns} it is expected to be in
the GeV to TeV range. The gravitino is a singlet with respect to the
gauge groups of the Standard Model. Its interactions---given by the
supergravity Lagrangian~\cite{Cremmer:1982en,Wess:1992cp}---are
suppressed by the (reduced) Planck scale~\cite{Yao:2006px}
\begin{align}
       \MPl=2.4\times 10^{18}\,\GeV \,.
\label{Eq:MPLmacro}
\end{align}
Once SUSY is broken, the extremely weak gravitino interactions are
enhanced through the super-Higgs mechanism, in particular, at
energy/mass scales that are large with respect to $\mgr$.
Nevertheless, the gravitino can be classified as an extremely weakly
interacting particle (EWIP). It must not be massive since even a light
gravitino can evade its production at colliders because of its tiny
interaction strength. Considering the case of the $\gravitino$ LSP, in
which the LOSP is the unstable NLSP that decays eventually into the
$\gravitino$ LSP, $\mst<\mneu$ (cf.~Fig.~\ref{Fig:YLOSP}) is viable as
already mentioned.%
\footnote{A stop $\widetilde{t}_1$ NLSP is not feasible in the
  CMSSM~\cite{DiazCruz:2007fc}.} 

% __________________________________________________________________
\subsection{Primordial Origin}
\label{sec:GravitinoProduction}
% __________________________________________________________________

Assuming that inflation governed the earliest moments of the Universe,
any initial population of gravitinos must be diluted away by the
exponential expansion during the slow-roll phase. Indeed, gravitinos
are typically not in thermal equilibrium with the primordial plasma
after inflation because of their extremely weak interactions.%
\footnote{In gauge-mediated SUSY breaking scenarios, light gravitinos
  can be viable thermal relics if their abundance is diluted by
  entropy production, which can result, for example, from decays of
  messenger
  fields~\cite{Baltz:2001rq,Fujii:2002fv,Fujii:2003iw,Lemoine:2005hu,Jedamzik:2005ir,Moultaka:2007pv}.}
At high temperatures, however, they can be produced efficiently in
thermal scattering of particles in the primordial plasma.
Derived in a consistent gauge-invariant treatment, the resulting
thermally produced (TP) gravitino density
reads~\cite{Bolz:2000fu,Pradler:2006qh}
\bea
        \Omega_{\gravitino}^{\TP}h^2
        &=&
        \sum_{i=1}^{3}
        \omega_i\, g_i^2 
        \left(1+\frac{M_i^2}{3\mgr^2}\right)
        \ln\left(\frac{k_i}{g_i}\right)
\nonumber\\
        &&
        \times
        \left(\frac{\mgr}{100~\GeV}\right)
        \left(\frac{T_{\Reheating}}{10^{10}\,\GeV}\right)
        \ ,
\label{Eq:GravitinoDensityTP}
\eea
with $\omega_i$, the gauge couplings $g_i$, the gaugino mass
parameters $M_i$, and $k_i$ as given in Table~\ref{Tab:Constants}. 
Here $M_i$ and $g_i$ are understood to be evaluated at the reheating
temperature%
\footnote{For a discussion on the $T_{\Reheating}$ definition, see
  Sec.~2 in~\cite{Pradler:2006hh}.}
after inflation $T_{\Reheating}$~\cite{Pradler:2006qh}.%
\footnote{Note that the field-theoretical methods applied in the
  derivation
  of~(\ref{Eq:GravitinoDensityTP})~\cite{Bolz:2000fu,Pradler:2006qh}
  require weak couplings $g_i\ll 1$ and thus $T \gg 10^6~\GeV$. For an
  alternative approach, see~\cite{Rychkov:2007uq}.}
\begin{table}[b]
   \caption{Assignments of the index $i$, the gauge coupling $g_i$, 
     and the gaugino mass parameter $M_i$ to the gauge groups
     U(1)$_\Hypercharge$, SU(2)$_\Weak$, and SU(3)$_\Color$,
     and the constants $\omega_i$ and $k_i$.}
   \label{Tab:Constants}
\begin{center}
\renewcommand{\arraystretch}{1.25}
\begin{tabular*}{3.25in}{@{\extracolsep\fill}ccccccc}
 \noalign{\smallskip}\hline
 gauge group         & $i$ & $g_i$ & $M_i$  &  $\omega_i$ &  $k_i$ 
 \\ 
 \noalign{\smallskip}\hline
 U(1)$_\Hypercharge$ & 1 & $g'$    & $M_1$  & 0.018 & 1.266  
 \\
 SU(2)$_\Weak$       & 2 & $g$     & $M_2$  & 0.044 & 1.312  
 \\
 SU(3)$_\Color$ & 3 & $g_\mathrm{s}$ & $M_3$ & 0.117 & 1.271  
 \\
 \noalign{\smallskip}\hline
\end{tabular*}
\end{center}
\end{table}
For the case of universal $M_{1,2,3}=\monetwo$ at $\mgut$ and
$\mgravitino \ll M_i$, i.e., $(1+M_i^2/3\mgr^2)\simeq M_i^2/3\mgr^2$,
$\Omegatp h^2$ can be approximated by the convenient
expression~\cite{Pradler:2007is}
\begin{align}
  \Omegatp h^2 \simeq 0.32 \Big( \frac{10\ \GeV}{\mgr} \Big) 
  \Big( \frac{\monetwo}{1\ \TeV} \Big)^2 \Big(
    \frac{\TR}{10^{8}\ \GeV} \Big) .
  \label{eq:omega-tp}
\end{align}
The thermally produced gravitinos do not affect the thermal evolution
of the LOSP (or NLSP) prior to its decay which occurs typically after
decoupling from the thermal plasma. Moreover, since each NLSP decays
into one $\gravitino$ LSP, the NLSP decay leads to a non-thermally
produced (NTP) gravitino
density~\cite{Borgani:1996ag,Asaka:2000zh,Feng:2003xh,Feng:2004mt}
\bea
        \Omega_{\gravitino}^{\NTP} h^2
        &=& 
        \mgravitino\, Y_{\NLSP}^{\dec}\, s(T_0) h^2 / \rho_{\mathrm{c}}
%        \ .
\label{Eq:GravitinoDensityNTP}
\eea
so that
the guaranteed density is given by%
\footnote{In this review I do not discuss gravitino production from
  inflaton decays which can be sub\-stan\-tial depending on the
  inflation model; see, e.g.,~\cite{Asaka:2006bv,Endo:2007sz}.}
 \begin{align}
   \Omega_{\gravitino}h^2=\Omega_{\gravitino}^{\TP}h^2+\Omega_{\gravitino}^{\NTP}h^2 
   \ .
   \label{Eq:GravitinoDensity}
 \end{align}

\begin{figure}[t!]
\includegraphics[width=.45\textwidth]{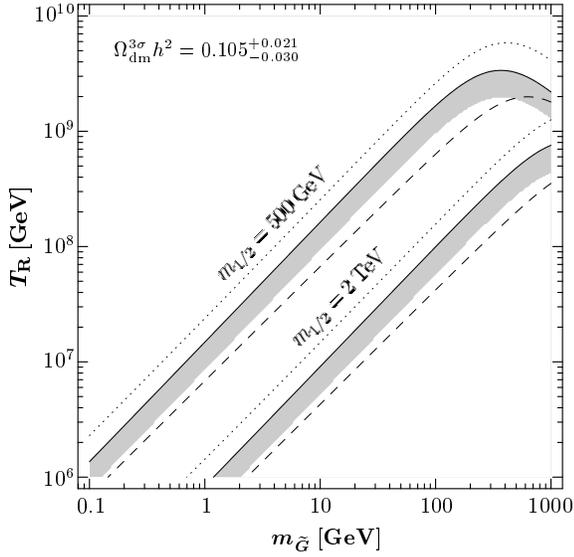} 
\caption{Upper limits on the reheating temperature $T_{\Reheating}$ in
  the $\gravitino$ LSP case.  On the upper (lower) gray band,
  $\Omega_{\widetilde{G}}^{\TP}\in\Omega_{\CDM}^{3\sigma}$ for
  $M_{1,2,3}=m_{1/2}=500~\GeV$ ($2~\TeV$) at $M_{\GUT}$. The
  corresponding limits from $\Omega_{\widetilde{G}}^{\TP}h^2\leq
  0.126$ shown by the dashed and dotted lines are obtained
  respectively with~(\ref{Eq:GravitinoDensityTP}) for
  $M_1/10=M_2/2=M_3=m_{1/2}$ at $M_{\GUT}$ and with the result of
  Ref.~\cite{Bolz:2000fu} for $M_3=m_{1/2}$ at $M_{\GUT}$.
  From~\cite{Pradler:2006hh}.}
\label{Fig:UpperLimitTR}
\end{figure}

\begin{figure}[t!]
  \includegraphics[width=0.45\textwidth]{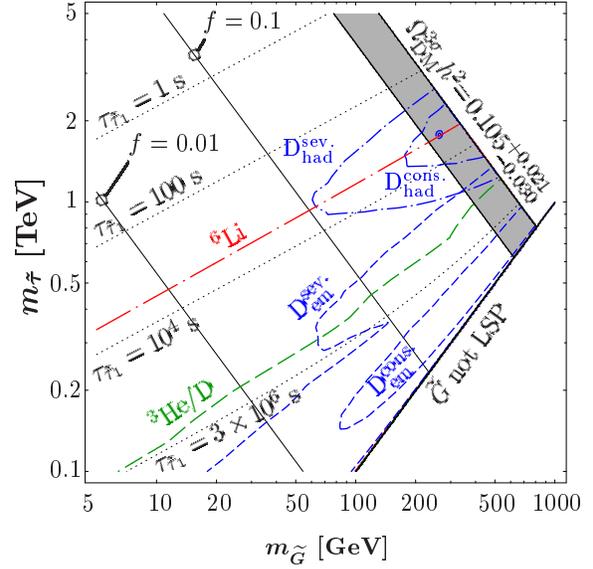}
  \caption{Cosmological constraints on the masses of the gravitino LSP
    and a purely `right-handed' $\stau$ NLSP. The gray band indicates
    $\Omega_{\gravitino}^{\NTP}\!\!\in\Omega_{\CDM}^{3\sigma}$. Above
    this band, $\Omega_{\gravitino} > 0.126$.  Only $10\%$ ($1\%$) of
    $\Omega_{\CDM}$ is provided by $\Omega_{\gravitino}^{\NTP}$ for
    scenarios that fall onto the thin solid line labeled by $f=0.1$
    ($0.01$). The dotted lines show contours of $\tau_{\stau}$.  The
    region below the long-dash-dotted (red in the web version) line
    and below the long-dashed (green in the web version) line is
    disfavored by the observationally inferred abundances of
    primordial $^6$Li~\cite{Pradler:2007is}
    and~$^3\mathrm{He}/\mathrm{D}$~\cite{Kawasaki:2004qu}. The effect
    of electromagnetic and hadronic energy injection on primordial D
    disfavors the regions inside the short-dash-dotted (blue in the
    web version) curves and to the right or inside of the short-dashed
    (blue in the web version) curves, respectively.
    With~(\ref{Eq:YstauNoCo}) and $\epsilon_{\EM}=0.3 E_{\tau}$, the
    curves are obtained from the severe and conservative upper limits
    defined in Sec.~4.1 of~\cite{Steffen:2006hw} based on results
    from~\cite{Cyburt:2002uv,Kawasaki:2004qu}. }
\label{Fig:GravitinoMassBounds}
\end{figure}

\begin{figure}[t!]
\includegraphics[width=.45\textwidth]{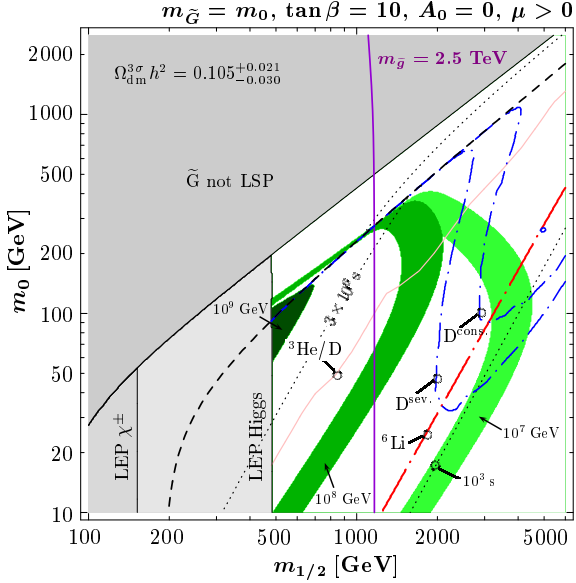} 
\caption{CMSSM regions with
  $\Omega_{\gravitino}h^2\in\Omega_{\CDM}^{3\sigma}$ for $\TR=10^7$,
  $10^8$, and $10^9~\GeV$ indicated respectively by the light, medium,
  and dark shaded (green in the web version) bands in the
  $(m_{1/2},m_0)$ planes for $\tan\beta=10$, $A_0=0$, $\mu>0$, and
  $\mgr=m_0$.  The regions excluded by the chargino and Higgs mass
  bounds and the line indicating $m_{\neutralino}=m_{\stau}$ are
  identical to the ones shown in Fig.~\ref{Fig:YLOSP}.  In the dark
  gray region, the gravitino is not the LSP. The dotted lines show
  contours of the NLSP lifetime.  The region to the left of the
  long-dash-dotted (red in the web version) line and to the left of
  the thin gray (pink in the web version) line is disfavored by the
  observationally inferred abundances of primordial
  $^6$Li~\cite{Pradler:2007is}
  and~$^3\mathrm{He}/\mathrm{D}$~\cite{Kawasaki:2004qu}.  The effect
  of hadronic energy injection on primordial D~\cite{Steffen:2006hw}
  disfavors the $\stau$ NLSP region above the short-dash-dotted (blue
  in the web version) lines.  The $\neutralino$ NLSP region is
  disfavored by BBN constraints from energy
  injection~\cite{Ellis:2003dn,Feng:2004mt,Roszkowski:2004jd,Cerdeno:2005eu,Cyburt:2006uv}.
  On the solid vertical line (violet in the web version)
  $m_{\widetilde{g}}=2.5\ \TeV$.  From~\cite{Pradler:2007ar}.}
\label{Fig:CMSSMtB10}
\end{figure}

While $\Omegatp$ is sensitive to $M_i$ and $\TR$ for a given
$\mgravitino$, $\Omegantp$ depends on
$Y_{\NLSP}^{\dec}=Y_{\LOSP}^{\dec}$ and thereby on details of the SUSY
model realized in nature; cf.\ Sect.~\ref{sec:NeutralinoProduction}.
For the case of the $\stau$ NLSP, simple approximations can be used
such as~\cite{Asaka:2000zh,Steffen:2006hw,Steffen:2006wx}
\begin{align}
        Y_{\stau}^{\dec}
        \simeq
        0.7 \times 10^{-12}
        \left(\frac{m_{\stau}}{1~\TeV}\right)
%        \ ,
\label{Eq:YstauNoCo}
\end{align}
which is valid outside of the $\stau$--$\neutralino$ coannihilation
region for a spectrum in which $m_{\stau}$ is significantly below the
masses of the lighter selectron and the lighter smuon, $m_{\stau} \ll
m_{\sel,\smu}$, and in which $\neutralino\simeq\Bino$ with a mass of
$m_{\Bino}=1.1\,m_{\stau}$.%
\footnote{The $Y_{\LOSP}^{\dec}$ contours in the $\stau$ LOSP region
  in Fig.~\ref{Fig:YLOSP}, in which $m_{\stau} \lesssim m_{\sel,\smu}
  \lesssim 1.1\,m_{\stau}$, illustrate that the $\stau$ LSP yield can
  be about twice as large for a given $m_{\stau}$ due to slepton
  coannihilation.  Approaching the $\neutralino$--$\stau$
  coannihilation region, $\mneu \approx \mst$, even larger factors
  occur.}
Scenarios with $\Omega_{\gravitino}=\Omega_{\CDM}$ are found for
natural mass spectra and for a wide range of $\mgr$--$\TR$
combinations.
This is illustrated in Figs.~\ref{Fig:UpperLimitTR},
\ref{Fig:GravitinoMassBounds}, and~\ref{Fig:CMSSMtB10}.

In $\gravitino$ LSP scenarios, upper limits on $T_{\Reheating}$ can be
derived since $\Omega_{\gravitino}^{\TP}\leq \Omega_{\CDM}$%
~\cite{Moroi:1993mb,Asaka:2000zh,Roszkowski:2004jd,Cerdeno:2005eu,Steffen:2006hw,Pradler:2006hh}.
These $\mgr$-dependent limits are shown in Fig.~\ref{Fig:UpperLimitTR}
(from~\cite{Pradler:2006hh}) and can be confronted with inflation
models.
Moreover, $T_{\Reheating}$ limits are important for our understanding
of the baryon asymmetry and, in particular, for thermal
leptogenesis~\cite{Fukugita:1986hr,Buchmuller:2004nz}.
%
% __________________________________________________________________
% __________________________________________________________________
%
For given $\Omega_{\gravitino}^{\TP}$, the bound
$\Omega_{\gravitino}^{\NTP}\leq\Omega_{\CDM}-\Omega_{\gravitino}^{\TP}$
gives upper limits on $\mgr$ and $\mst$. The limits obtained for a
$\stau$ NLSP with~(\ref{Eq:YstauNoCo}) are shown in
Fig.~\ref{Fig:GravitinoMassBounds}.
%
% __________________________________________________________________
% __________________________________________________________________
%
In Fig.~\ref{Fig:CMSSMtB10} (from~\cite{Pradler:2007ar}) regions with
$\Omega_{\gravitino}\in\Omega_{\CDM}^{3\sigma}$ are shown for
$\TR=10^7$, $10^8$, and $10^9~\GeV$. Here both
$\Omega_{\gravitino}^{\TP}$ and $\Omega_{\gravitino}^{\NTP}$ are taken
into account for~$\mgr=m_0$ within the framework of the CMSSM.
%
% __________________________________________________________________
% __________________________________________________________________

While thermally produced gravitinos have a negligible free--streaming
velocity today, gravitinos from NLSP decays can be warm/hot dark
matter.  In the $\stau$ NLSP case, for example, upper limits on the
free--streaming velocity from simulations and observations of cosmic
structures exclude
$\mst\lesssim 0.7~\TeV$
for
$\Omega_{\gravitino}^{\NTP}\!\!\simeq\Omega_{\CDM}$~\cite{Steffen:2006hw}.
Such scenarios (gray band in Fig.~\ref{Fig:GravitinoMassBounds}),
however, require $\mst\gtrsim 0.7~\TeV$ anyhow and could even resolve
the small scale structure problems inherent to cold dark
matter~\cite{Cembranos:2005us,Kaplinghat:2005sy,Jedamzik:2005sx}.

% __________________________________________________________________
\subsection{Cosmological Constraints}
\label{sec:GravitinoConstraints}
% __________________________________________________________________

In the $\gravitino$ LSP case with conserved R-parity, the NLSP can
have a long lifetime $\tau_{\NLSP}$.%
\footnote{For the case of broken R-parity, see, e.g.,~\cite{Takayama:2000uz,Buchmuller:2007ui,Ibarra:2007jz}.}
This is illustrated by the dotted $\tau_{\NLSP}$ contours in
Figs.~\ref{Fig:GravitinoMassBounds} and~\ref{Fig:CMSSMtB10}.
In particular, for the $\stau$ NLSP, one finds in the limit
$m_{\tau}\to 0$,
\begin{equation}
        \tau_{\st} 
        \simeq 
        \Gamma^{-1}(\stau\to\gravitino\tau)
        = 
        \frac{48 \pi \mgr^2 \MPl^2}{\mst^5}\!
        \left(\!1-\frac{\mgr^2}{\mst^2}\right)^{\!\!\!\!-4},
\label{Eq:StauLifetime}
\end{equation}
while the expression for the lifetime of the $\neutralino$ NLSP is
given in Sec.~IIC of Ref.~\cite{Feng:2004mt}.

If the NLSP decays into the $\gravitino$ LSP occur during or after
big-bang nucleosynthesis (BBN), the Standard Model particles emitted
in addition to the gravitino can affect the abundances of the
primordial light elements. Indeed, these BBN constraints disfavor the
$\neutralino$ NLSP for $\mgr\gtrsim
100~\MeV$~\cite{Feng:2004mt,Roszkowski:2004jd,Cerdeno:2005eu,Cyburt:2006uv}.
For the slepton NLSP case, the BBN constraints associated with
electromagnetic/hadronic energy injection have also been considered
and found to be much weaker but still significant in much of the
parameter
space~\cite{Feng:2004mt,Roszkowski:2004jd,Cerdeno:2005eu,Steffen:2006hw}
as can be seen in Fig.~\ref{Fig:GravitinoMassBounds}, where the
constraints from electromagnetic and hadronic energy release are shown
respectively by the short-dashed and the short-dash-dotted (blue in
the web version) lines. The hadronic constraints are also shown in
Fig.~\ref{Fig:CMSSMtB10}.

It has been realized only recently that already the mere presence of
long-lived negatively charged particles can affect BBN substantially
via bou\-nd-state effects%
~\cite{Pospelov:2006sc,Kohri:2006cn,Kaplinghat:2006qr,Cyburt:2006uv,Hamaguchi:2007mp,Bird:2007ge,Kawasaki:2007xb,Jittoh:2007fr,Jedamzik:2007cp}.
In particular, bou\-nd-state formation of $\stau^-$ with $^4$He can
lead to an overproduction of $^6$Li via the catalyzed BBN (CBBN)
reaction
$(\Hefour\champ)+\mathrm{D} \rightarrow \Lisix + \champ$~\cite{Pospelov:2006sc}.
Thereby, the observationally inferred upper limit on the primordial
$\Lisix$ abundance
$\Lisix/\mathrm{H} |_{\mathrm{obs}} \lesssim 2\times 10^{-11}$ \cite{Cyburt:2002uv},
leads to the CBBN constraint shown by the long-dash-dotted (red in the
web version) lines in Figs.~\ref{Fig:GravitinoMassBounds}
and~\ref{Fig:CMSSMtB10}, as obtained from~\cite{Pradler:2007is}.
Indeed, for a typical yield~(\ref{Eq:YstauNoCo}), the
$\Lisix/\mathrm{H} |_{\mathrm{obs}}$ limit quoted above implies the
constraint~\cite{Pospelov:2006sc,Hamaguchi:2007mp,%
  Bird:2007ge,Takayama:2007du,Pradler:2007is}: $\taustau \lesssim
5\times 10^3\;\seconds$.
While numerous other CBBN reactions can affect the abundances of
$\Lisix$ and other primordial elements
significantly~\cite{Cyburt:2006uv,Bird:2007ge,Jedamzik:2007cp,Jedamzik:2007qk},
the approximate $\taustau$ bound is relatively robust.  In fact, by
systematically taking into account the uncertainties in the relevant
nuclear reaction rates, it is shown explicitly in Fig.~14
of~\cite{Jedamzik:2007cp} and in Fig.~5 of~\cite{Jedamzik:2007qk} that
cosmologically allowed regions for $\taustau \gtrsim 10^5\;\seconds$
are extremely unlikely.  In particular, the $^3$He/D constraint on
electromagnetic energy release~\cite{Sigl:1995kk} becomes severe and
can exclude $\taustau\gtrsim
10^6~\seconds$~\cite{Cerdeno:2005eu,Cyburt:2006uv,Kawasaki:2007xb,Jedamzik:2007cp}.
This is shown by the long-dashed (green in the web version) line in
Fig.~\ref{Fig:GravitinoMassBounds} and by the thin gray (pink in the
web version) line in Fig.~\ref{Fig:CMSSMtB10}, which are obtained from
Fig.~42 of Ref.~\cite{Kawasaki:2004qu} for a `visible' electromagnetic
energy of $E_{\mathrm{vis}}=\epsilon_{\EM}= 0.3\,E_{\tau}$ of the tau
energy $E_{\tau}=(\mst^2-\mgr^2+m_{\tau}^2)/2\mst$ released in
$\stau\to\gravitino\tau$.%
\footnote{With a finely tuned $\mst$--$\mgr$ degeneracy leading to
  $E_{\mathrm{vis}}\to 0$ can any bound on energy release be evaded.}

The observed Planck spectrum of the cosmic microwave background (CMB)
provides an additional constraint~\cite{Hu:1992dc,Lamon:2005jc} which
is not shown. Indeed, the CMB limit derived in~\cite{Lamon:2005jc} is
everywhere less severe than the severe electromagnetic limit
$D_{\EM}^{\mathrm{sev.}}$ given by the short-dashed (blue in the web
version) line in Fig.~\ref{Fig:GravitinoMassBounds}.

As can be seen in Fig.~\ref{Fig:GravitinoMassBounds}, the cosmological
constraints provide an upper bound on $\mgravitino$ once $m_{\st}$ is
measured. This bound implies upper bounds on the SUSY breaking scale,
$\Omega_{\gravitino}^{\NTP}$, and $\TR$.

Figure~\ref{Fig:CMSSMtB10} shows that the cosmological constraints
imply a lower limit on $\monetwo$~\cite{Cyburt:2006uv,Pradler:2006hh}
and an upper limit on $\TR$~\cite{Pradler:2006hh}.  Indeed, from
$\taustau \lesssim 5\times 10^3\;\seconds$, $\mgr$-dependent limits on
the gaugino mass parameter,
 \begin{align}
  \label{eq:LowerLimitm12}
  \monetwo &\ge 0.9\, \TeV \left( \frac{\mgr}{ 10\ \GeV}
  \right)^{2/5},
\end{align}
and the reheating temperature,
\begin{align}
    \label{eq:UpperLimitTR}
  \TR &\le 4.9\times 10^7 \ \GeV \left( \frac{\mgr}{10\ \GeV}
  \right)^{1/5},
\end{align}
have been derived within the CMSSM~\cite{Pradler:2007is}.%
\footnote{Similar limits have recently been discussed in models where
  the ratio $\mgr/\monetwo$ is bounded from
  below~\cite{Kersten:2007ab}.}
While the $\TR$ bound can be restrictive for models of inflation and
baryogenesis, the $\monetwo$ bound can have implications for SUSY
searches at the LHC. Depending on $\mgr$, (\ref{eq:LowerLimitm12})
implies sparticle masses which can be associated with a mass range
that will be difficult to probe at the LHC.  This is illustrated by
the vertical (violet in the web version) line in
Fig.~\ref{Fig:CMSSMtB10} which indicates the gluino mass
$m_{\widetilde{g}} = 2.5\ \TeV$~\cite{Pradler:2007ar}.%
\footnote{Note that the mass of the lighter stop is
  $m_{\widetilde{t}_1} \simeq 0.7 m_{\widetilde{g}}$ in the considered
  $\stau$ NLSP region with $m_{\mathrm{h}}>114.4\ \GeV$.}
%

% __________________________________________________________________
\subsection{Experimental Prospects}
\label{sec:GravitinoExperiments}
% __________________________________________________________________

Because of its extremely weak couplings, gravitino dark matter is
inaccessible to direct and indirect searches if R-parity is
conserved.%
\footnote{For broken R-parity, $\gravitino$ dark matter is unstable so
  that decay products can appear in indirect
  searches~\cite{Buchmuller:2007ui,Bertone:2007aw,Ibarra:2007wg}.}
Also the direct production of gravitinos at colliders is strongly
suppressed.  Instead, one expects a large sample of (quasi-) stable
NLSPs if the NLSP belongs to the MSSM spectrum. 

In the $\stau$ NLSP case, each heavier superpartner produced will
cascade down to the $\stau$ which will appear as a (quasi-) stable
particle in the detector. Such a heavy charged particle would
penetrate the collider detector in a way similar to
muons~\cite{Drees:1990yw,Nisati:1997gb,Feng:1997zr}.  If the produced
staus are slow, the associated highly ionizing tracks and
time--of--flight measurements will allow one to distinguish the
$\stau$ from a
muon~\cite{Drees:1990yw,Nisati:1997gb,Feng:1997zr,Ambrosanio:2000ik}.
With measurements of the $\stau$ velocity $\beta_{\stau} \equiv
v_{\stau}/c$ and the slepton momentum $p_{\stau}\equiv
|\vec{p}_{\stau}|$, $\mst$ can be determined:
$\mst=p_{\stau}(1-\beta_{\stau}^2)^{1/2}/\beta_{\stau}$~\cite{Ambrosanio:2000ik}.
For the upcoming LHC experiments, studies of hypothetical scenarios
with long-lived charged particles are actively
pursued~\cite{Ellis:2006vu,Ellis:2007mc,Bressler:2007gk,Zalewski:2007up}.
For example, it has been found that one should be able to measure the
mass $\mst$ of a (quasi-) stable $\stau$ quite
accurately~\cite{Ellis:2006vu,Ellis:2007mc}.%
\footnote{(Quasi-) stable $\stau$'s could also be pair-produced in
  interactions of cosmic neutrinos in the earth matter and be detected
  in a neutrino telescope such as IceCube~\cite{Ahlers:2006pf}.}

If some of the staus decay already in the collider detectors, the
statistical method proposed in~\cite{Ambrosanio:2000ik} could allow
one to measure the $\stau$ lifetime. With~(\ref{Eq:StauLifetime}) and
the measured value of $m_{\stau}$, one will then be able to determine
also the gravitino mass $\mgravitino$ and thereby the scale of SUSY
breaking. As a test of our understanding of the early Universe, it
will also be interesting to confront the experimentally determined
$(\mgravitino,m_{\stau})$ point with the cosmological constraints in
Fig.~\ref{Fig:GravitinoMassBounds}.

Ways to stop and collect charged long-lived particles for an analysis
of their decays have also been
proposed~\cite{Goity:1993ih,Hamaguchi:2004df,Feng:2004yi,DeRoeck:2005bw,Hamaguchi:2006vu,Cakir:2007xa}.
It was found that up to $\Order(10^3$--$10^4)$ and
$\Order(10^3$--$10^5)$ $\stau$'s can be trapped per year at the LHC
and the ILC, respectively, by placing 1--10~kt of massive additional
material around planned collider
detectors~\cite{Hamaguchi:2004df,Feng:2004yi}. A measurement of
$\tau_{\stau}$ can then be used to determine $\mgravitino$ as already
described above.
If $\mgr$ can be determined independently from the kinematics of the
2-body decay $\stau\to\gravitino\tau$,
\begin{align}
      \mgr = \sqrt{{m_{\st}^2}+{m_\tau^2}-2{m_{\st} E_\tau}}  
      \, ,
\label{Eq:mgrKinematics}
\end{align}
the lifetime $\tau_{\st}$ can allow for a measurement of the Planck
scale~\cite{Buchmuller:2004rq,Martyn:2006as,Hamaguchi:2006vu,Martyn:2007mj}
\begin{align}
  \MPl^2 =  
  \frac{\tau_{\stau}}{48\pi} 
  \frac{\mst^5}{\mgr^2}
  \left(
  1 - \frac{\mgr^2}{\mst^2}
  \right)^4
  .
\label{Eq:Planck_Scale}
\end{align}
An agreement with~(\ref{Eq:MPLmacro}), which is inferred from Newton's
constant~\cite{Yao:2006px}
$G_{\rm N} = 6.709\times 10^{-39}\,\GeV^{-2}$,
would then provide strong evidence for the existence of supergravity
in nature~\cite{Buchmuller:2004rq}. In fact, this agreement would be a
striking signature of the gravitino LSP. Unfortunately, the required
kinematical determination of $\mgravitino$ appears to be feasible only
for~\cite{Martyn:2006as,Hamaguchi:2006vu,Martyn:2007mj}
$\mgravitino/\mst \gtrsim 0.1$
which seems to be disfavored according to our present understanding of
the cosmological constraints (see
Fig.~\ref{Fig:GravitinoMassBounds}).%
\footnote{Note that the cosmological constraints described in
  Sect.~\ref{sec:GravitinoConstraints} assume a standard thermal
  history. In fact, entropy production after NLSP decoupling and
  before BBN can weaken the BBN constraints
  significantly~\cite{Buchmuller:2006tt,Pradler:2006hh}.}
Accordingly, alternative methods such as the ones proposed
in~\cite{Brandenburg:2005he,Steffen:2005cn} could become essential to
identify the gravitino as the LSP.

% __________________________________________________________________
\section{Axino Dark Matter}
\label{sec:AxinoDM}
% __________________________________________________________________

The axino
$\axino$~\cite{Nilles:1981py,Kim:1983dt,Tamvakis:1982mw,Kim:1983ia}
appears (as the spin-1/2 superpartner of the axion) once the MSSM is
extended with the Peccei--Quinn
mechanism~\cite{Peccei:1977hh,Peccei:1977ur} in order to solve the
strong CP problem. Depending on the model and the SUSY breaking
scheme, the axino mass $m_{\axino}$ can range between the eV and the
GeV
scale~\cite{Tamvakis:1982mw,Nieves:1985fq,Rajagopal:1990yx,Goto:1991gq,Chun:1992zk,Chun:1995hc}.
The axino is a singlet with respect to the gauge groups of the
Standard Model. It interacts extremely weakly since its couplings are
suppressed by the Peccei--Quinn
scale~\cite{Yao:2006px,Sikivie:2006ni,Raffelt:2006rj,Raffelt:2006cw}
\begin{align}
        f_a\gtrsim 5\times 10^9\,\GeV   
\label{Eq:fPQ}
\end{align}
and thus can be classified as an EWIP.  The detailed form of the axino
interactions depends on the axion model under consideration.  We focus
on hadronic (or KSVZ) axion models~\cite{Kim:1979if,Shifman:1979if} in
a SUSY setting, in which the axino couples to the MSSM particles only
indirectly through loops of additional heavy KSVZ (s)quarks.
Considering $\axino$ LSP scenarios in which the LOSP is the NLSP,
$\mst<\mneu$ is again viable.

Before proceeding, it should be stressed that the bosonic partners of
the axino, the axion and the saxion, can have important implications
for cosmology: (i)~The relic density of axions can contribute
significantly to the dark matter
density~\cite{Sikivie:2006ni,Raffelt:2006rj,Wilczek:2007gsa} and
thereby tighten the constraints from $\Omega_{\axino}^{\TP} <
\Omega_{\CDM}$ discussed below. (ii)~Late decays of the saxion can
lead to significant entropy
production~\cite{Kim:1992eu,Lyth:1993zw,Chang:1996ih,Hashimoto:1998ua}
and thereby affect the cosmological
constraints~\cite{Kawasaki:2007mk}. In this review, however, a
standard thermal history is assumed which implies that saxion effects
are negligible.

% __________________________________________________________________
\subsection{Primordial Origin}
\label{sec:AxinoProduction}
% __________________________________________________________________

Because of their extremely weak interactions, the temperature
$T_{\freezeout}$ at which axinos decouple from the thermal plasma in
the early Universe is very high. For example, an axino decoupling
temperature of $T_{\freezeout} \approx 10^9\,\GeV$ is obtained for
$f_a=10^{11}\,\GeV$~\cite{Rajagopal:1990yx,Brandenburg:2004du}.  For
$\TR>T_{\freezeout}$, axinos were in thermal equilibrium before
decoupling as a relativistic species so
that~\cite{Rajagopal:1990yx,Asaka:2000ew,Covi:2001nw,Brandenburg:2004du}
\begin{align}
  \Omega_{\ax}^{\equil}h^2
  \approx\frac{m_{\ax}}{2\,\keV}
  \ .
  \label{Eq:AxinoDensityEq}
\end{align}
For $\TR<T_{\freezeout}$, axinos are not in thermal equilibrium with
the primordial plasma after inflation but can be generated efficiently
in scattering processes of particles that are in thermal equilibrium
with in the hot SUSY
plasma~\cite{Asaka:2000ew,Covi:2001nw,Brandenburg:2004du}.  Within
SUSY QCD, the associated thermally produced (TP) axino density can be
calculated in a consistent gauge-invariant treatment that requires
weak couplings ($g_\mathrm{s}\ll 1$)~\cite{Brandenburg:2004du}:
\bea
        \Omega_{\axino}^{\TP}h^2
        &\simeq&
        5.5\,g_\mathrm{s}^6 \ln\left(\frac{1.108}{g_\mathrm{s}}\right) 
        \left(\frac{10^{11}\,\GeV}{f_a/N}\right)^{\! 2}\!\!
\nonumber\\
        &&
        \times
        \bigg(\frac{m_{\ax}}{0.1~\GeV}\bigg)\!
        \left(\frac{T_R}{10^4\,\GeV}\right)
%        \ ,
\label{Eq:AxinoDensityTP}
\eea
with the axion-model-dependent color anomaly $N$ of the Peccei--Quinn
symmetry and the strong coupling $g_\mathrm{s}$ understood to be
evaluated at $\TR$. The thermally produced axinos do not affect the
thermal evolution of the LOSP (or NLSP) which decays after its
decoupling into the $\axino$ LSP. Taking into account the
non-thermally produced (NTP) density from NLSP
decays~\cite{Covi:1999ty,Covi:2001nw}
\bea
        \Omega_{\axino}^{\NTP} h^2
        &=& 
        m_{\axino}\, Y_{\NLSP}^{\dec}\, s(T_0) h^2 / \rho_{\mathrm{c}}
        \ ,
\label{Eq:AxinoDensityNTP}
\eea
the guaranteed axino density is%
\footnote{Axino production in inflaton decays is not considered.}
\begin{align}
\Omega_{\axino}=\Omega_{\axino}^{\,\,\equil/\TP}+\Omega_{\axino}^{\NTP}
\ .
\end{align}

\begin{figure}[t]
\includegraphics*[width=.45\textwidth]{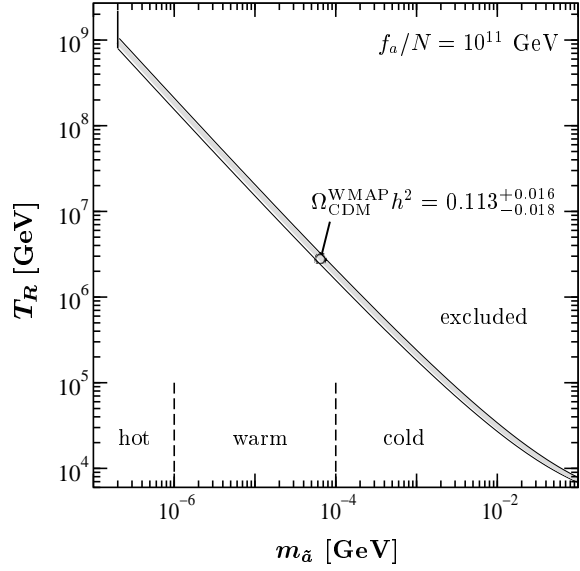}
\caption{Upper limits on the reheating temperature $T_{\Reheating}$ in
  the $\axino$ LSP case for $f_a/N = 10^{11}\,\GeV$. On (above) the
  gray band, $\Omega^{\TP}_{\ax}h^2\in 0.113^{+0.016}_{-0.018}$
  ($\Omega^{\TP}_{\ax}h^2> 0.129$).  Thermally produced axinos can be
  classified as hot, warm, and cold dark matter~\cite{Covi:2001nw} as
  indicated.  From~\cite{Brandenburg:2004du}.}
\label{Fig:axino_dark_matter_limits}
\end{figure}

In Fig.~\ref{Fig:axino_dark_matter_limits}
(from~\cite{Brandenburg:2004du}) the $(m_{\ax},\TR)$ region with
$0.097\leq\Omega_{\axino}^{\TP}\leq 0.129$ for $f_a/N = 10^{11}\,\GeV$
is shown by the gray band.  Note that~(\ref{Eq:AxinoDensityTP}) shows
a different dependence on the LSP mass than the corresponding
expression in the~$\gravitino$ LSP case~(\ref{Eq:GravitinoDensityTP}).
Accordingly, one finds the different $m_{\LSP}$ dependence of the
$T_{\Reheating}$ limits inferred from
$\Omega_{\axino/\gravitino}^{\TP} < \Omega_{\CDM}$.
%
% __________________________________________________________________
% __________________________________________________________________
%
Since thermally produced axinos are generated in kinetic equilibrium
with the primordial plasma, they have a thermal spectrum which allows
for the $m_{\ax}$-dependent classification into cold, warm, and hot
dark matter~\cite{Covi:2001nw} shown in
Fig.~\ref{Fig:axino_dark_matter_limits}.
As can be seen, the $\TR$ limit does not exist for $m_{\ax}\lesssim
0.2~\keV$ because of the equality of $\axino$ production and $\axino$
disappearance rates for $T > T_{\freezeout}\approx 10^9\,\GeV$.  With
a thermal relic density~(\ref{Eq:AxinoDensityEq}) in this regime,
there will be a limit on $m_{\ax}$ depending on the constraints
inferred from studies of warm/hot dark matter~\cite{Hannestad:2007dd}.

The non-thermally produced axino density $\Omega_{\axino}^{\NTP}$
differs from the corresponding expression in the $\gravitino$ LSP
case~(\ref{Eq:GravitinoDensityNTP}) only by the obvious difference in
$m_{\axino/\gravitino}$. In particular, for given
$\Omega_{\axino}^{\TP}$, the bound
$\Omega_{\axino}^{\NTP}\leq\Omega_{\CDM}-\Omega_{\axino}^{\,\equil/\TP}$ 
as obtained with~(\ref{Eq:YstauNoCo}) implies limits on $m_{\ax}$ and
$\mst$ which can be read off directly from
Fig.~\ref{Fig:GravitinoMassBounds} after the replacement
$\mgr\rightarrow m_{\ax}$.  Note, however, that the $\taustau$
contours and the cosmological constraints are different in the axino
LSP case. For the $\stau$ NLSP, the following lifetime was
estimated~\cite{Brandenburg:2005he}
\bea
        \taustau 
        && \!\!\simeq 
        \Gamma^{-1}(\stau\to\tau\,\axino)
        \simeq
        25~\mathrm{s}\,\,  
        \xi^{-2}\,
        \left(1-\frac{m_{\ax}^2}{m_{\st}^2}\right)^{\!\!\!\!-1}
\nonumber\\
        &&
        \times
        \left(\frac{100\,\GeV}{m_{\st}}\right)\!\!
        \left(\frac{f_a/C_{\rm aYY}}{10^{11}\,\GeV}\right)^{\!\!2}\!\!
        \left(\frac{100\,\GeV}{m_{\tilde{B}}}\right)^{\!\!2}\!,
\label{Eq:Axino2Body}
\eea
where the KSVZ-model dependence is expressed by $C_{\rm aYY}\simeq
\mathcal{O}(1)$ and the uncertainty of the estimate is absorbed into
$\xi\simeq\Order(1)$. One thus finds a $\stau$ lifetime in the
$\axino$ LSP case that cannot be as large as the one in the
$\gravitino$ LSP case~(\ref{Eq:StauLifetime}). Accordingly, the BBN
constraints are much weaker for the $\axino$ LSP. For discussions of
$\axino$ LSP scenarios with the $\neutralino$ NLSP,
see~\cite{Covi:1999ty,Covi:2001nw,Covi:2004rb}. For both the $\stau$
NLSP and the $\neutralino$ NLSP, it has been shown that non-thermally
produced axinos with $m_{\ax}\lesssim 10~\GeV$ would be warm/hot dark
matter~\cite{Jedamzik:2005sx}.

% __________________________________________________________________
\subsection{Experimental Prospects}
\label{sec:AxinoExperiments}
% __________________________________________________________________

Being an EWIP, the axino LSP is inaccessible to any direct and
indirect dark matter searches if R-parity is conserved. Also the
direct $\axino$ production at colliders is strongly suppressed.
Nevertheless, (quasi-) stable $\stau$'s could appear in collider
detectors (and neutrino telescopes~\cite{Ahlers:2006pf}) as a possible
signature of the $\axino$ LSP. However, since the $\MPl$ measurement
at colliders~\cite{Buchmuller:2004rq}, which would have been a
decisive test of the $\gravitino$ LSP, seems cosmologically
disfavored, it will be a challenge to distinguish between the $\axino$
LSP and the $\gravitino$ LSP.

For $m_{\st}=100\,\GeV$ and $m_{\Bino}=110\,\GeV$, for example, the
$\stau$ lifetime in the $\axino$ LSP scenario~(\ref{Eq:Axino2Body})
can range from $\Order(0.01~{\mbox{s}})$ for $f_a=5\times 10^9\,\GeV$
to $\Order(10~{\mbox{h}})$ for $f_a=5\times 10^{12}\,\GeV$. In the
$\gravitino$ LSP case, the corresponding
lifetime~(\ref{Eq:StauLifetime}) can vary over an even wider range,
e.g., from $6\times 10^{-8}\,{\rm s}$ for $\mgravitino = 1~\keV$ to
15~years for $\mgravitino = 50~\GeV$. Thus, both a very short
lifetime, $\taustau \lesssim$~ms, and a very long one, $\taustau
\gtrsim$~days, will point to the $\gravitino$ LSP. On the other hand,
if the LSP mass cannot be measured kinematically and if
$\taustau=\Order(0.01~{\mbox{s}})$--$\Order(10~{\mbox{h}})$, the stau
lifetime alone will not allow us to distinguish between the $\axino$
LSP and the $\gravitino$ LSP.

\begin{figure}[t]
\includegraphics*[width=0.45\textwidth]{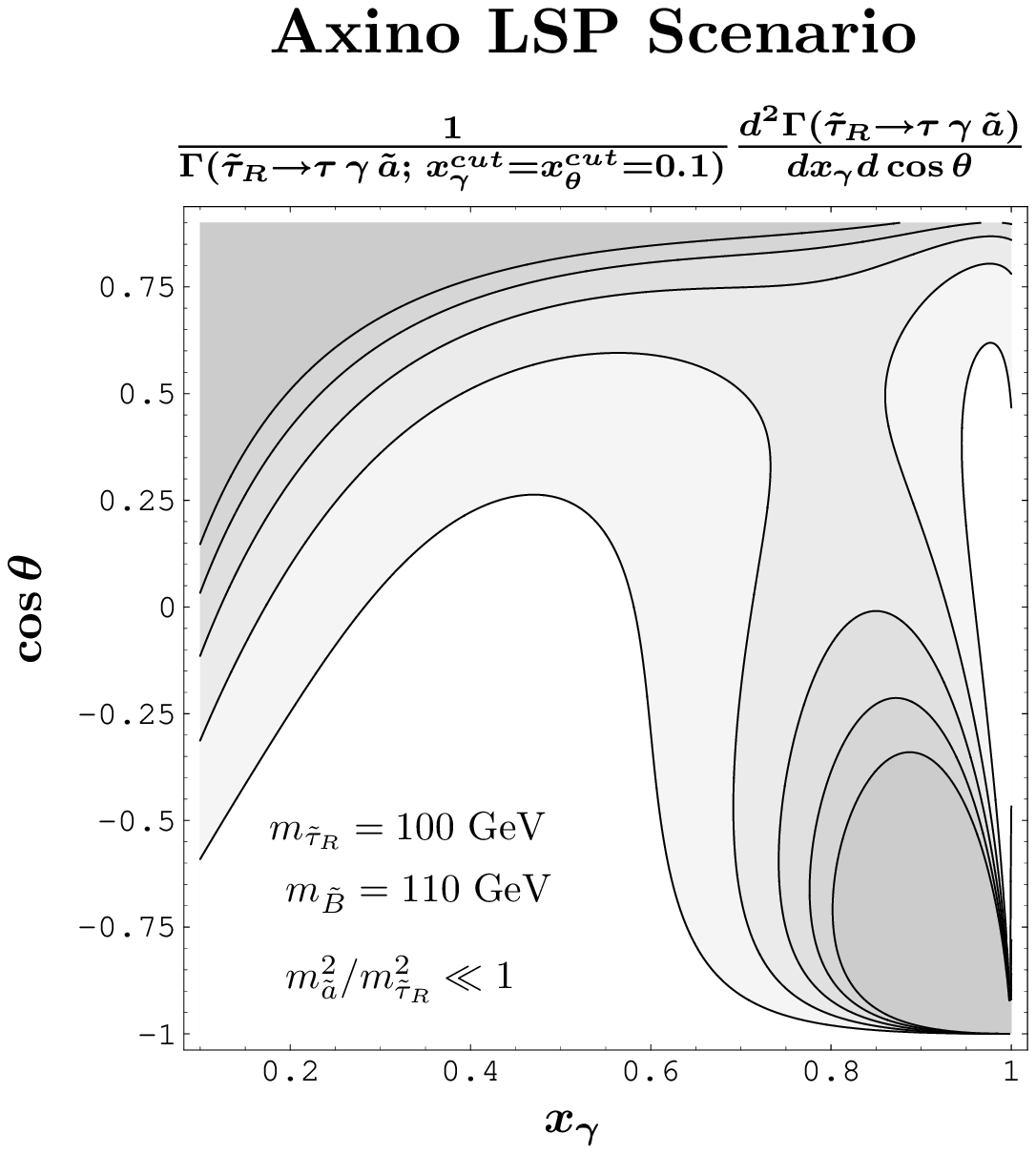}\\ \\
\includegraphics*[width=0.45\textwidth]{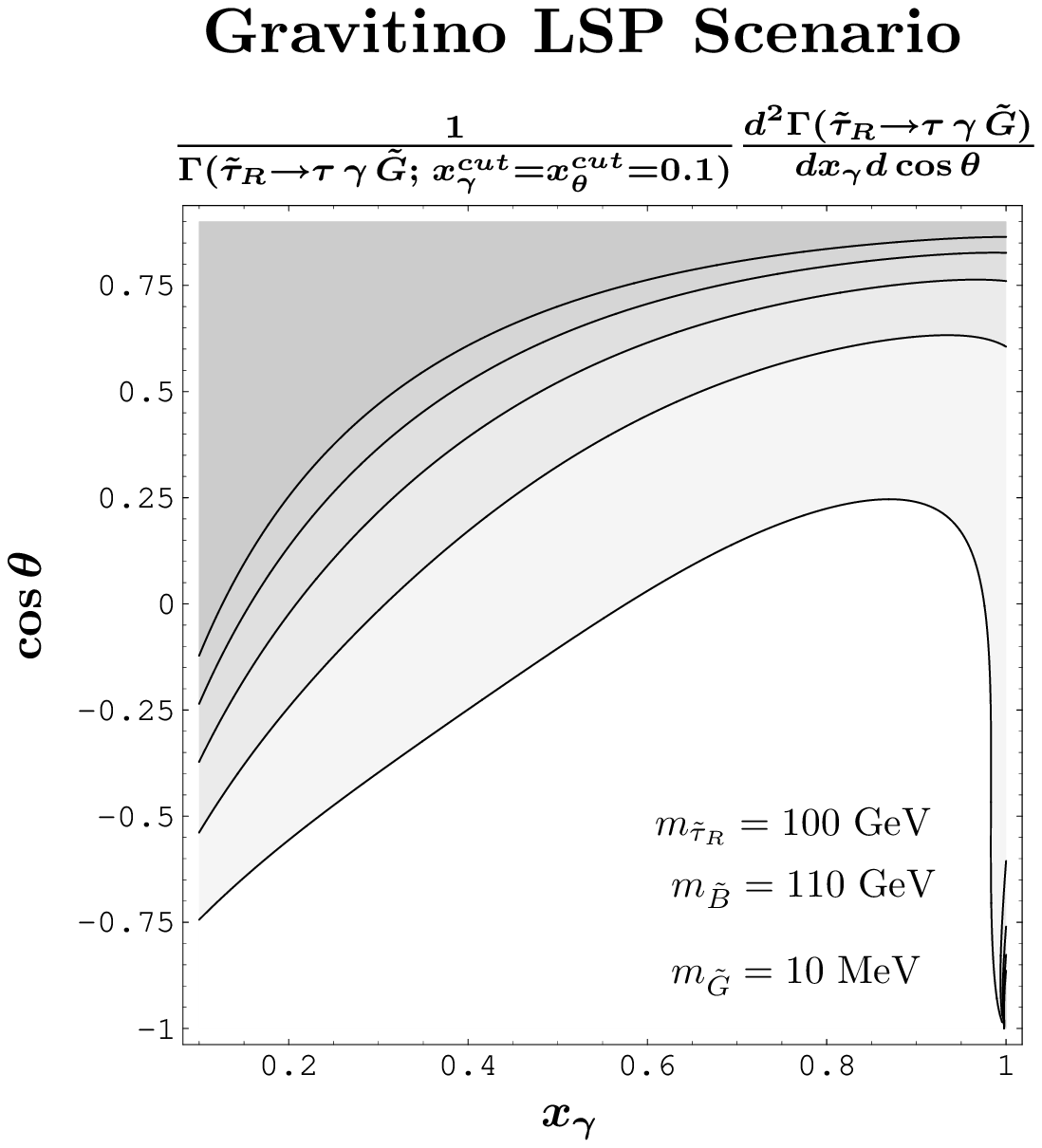}
\caption{ The normalized differential
  distributions~(\ref{Eq:Fingerprint}) of the visible decay products
  in the decays $\stau\to\tau+\gamma+\axino/\gravitino$ for the cases
  of the $\axino$ LSP (upper panel) and the $\gravitino$ LSP (lower
  panel) for $m_\stau = 100\,\GeV$, $\neutralino\simeq\Bino$, $m_\Bino
  = 110\,\GeV$, $m_{\axino}^2/m_{\stau}^2 \ll 1$, and $\mgravitino =
  10\,\MeV$.  The cut parameters are set to $x_\gamma^{\mathrm{cut}} =
  x_\theta^{\mathrm{cut}}=0.1$. The contour lines represent the values
  0.2, 0.4, 0.6, 0.8, and 1.0, where the darker shading implies a
  higher number of events. From~\cite{Brandenburg:2005he}.}
\label{Fig:Fingerprint}
\end{figure}

The situation is considerably improved when one considers the 3-body
decays $\stau\to \tau \gamma \axino/\gravitino$. From the
corresponding differential rates~\cite{Brandenburg:2005he}, one
obtains the differential distributions of the visible decay products.
These are illustrated in Fig.~\ref{Fig:Fingerprint}
(from~\cite{Brandenburg:2005he}) in terms of
\begin{align}
        {1 
        \over 
        \Gamma(\stau\to\tau\,\gamma\, i\,;
        x_{\gamma}^{\mathrm{cut}},x_{\theta}^{\mathrm{cut}})}
        \,\,
        {d^2\Gamma(\stau\to\tau\,\gamma\, i)
        \over
        d x_{\gamma}d \cos\theta}
%        \ ,
\label{Eq:Fingerprint}
\end{align}
where $x_\gamma\equiv 2 E_\gamma/m_{\st}$ is the scaled photon energy,
$\theta$ the opening angle between the directions of $\gamma$ and
$\tau$,
\bea
        \Gamma(\stau\to\tau\,\gamma\,i\,;
        x_\gamma^{\mathrm{cut}},x_\theta^{\mathrm{cut}})
        &\equiv&
        \int^{1-A_{i}}_{x_\gamma^{\mathrm{cut}}}
        d x_\gamma
        \int^{1-x_\theta^{\mathrm{cut}}}_{-1}
        d \cos\theta \,\,
\nonumber\\
        &&\times
        \frac{d^2\Gamma(\stau\to\tau\,\gamma\,i)}{dx_\gamma d\cos\theta}
\eea
the respective integrated 3-body decay rate with the cuts $x_\gamma >
x_\gamma^{\mathrm{cut}}$ and $\cos\theta < 1-x_\theta^{\mathrm{cut}}$,
and $A_{i} \equiv m_{i}^2/m_{\st}^2$.
Note that~(\ref{Eq:Fingerprint}) is independent of the 2-body decay,
the total NLSP decay rate, and the Peccei--Quinn/Planck scale.
%
% __________________________________________________________________
% __________________________________________________________________
%
The figure shows~(\ref{Eq:Fingerprint}) for the axino LSP ($i =
\axino$) with $m_{\ax}^2/m_{\stau}^2 \ll 1$ (upper panel) and the
gravitino LSP ($i=\gravitino$) with $\mgravitino = 10~\MeV$ (lower
panel), where $m_{\st} = 100~\GeV$, $m_{\Bino} = 110~\GeV$, and
$x_\gamma^{\mathrm{cut}} = x_\theta^{\mathrm{cut}}=0.1$.
In the $\gravitino$ LSP case, the events are peaked only in the region
where photons are soft and emitted with a small opening angle with
respect to the tau ($\theta\simeq 0$).  In contrast, in the $\axino$
LSP case, the events are also peaked in the region where the photon
energy is large and the photon and the tau are emitted back-to-back
($\theta \simeq \pi$).
Thus, if the observed number of events peaks in both regions, this can
be evidence against the gravitino LSP and a hint towards the axino
LSP~\cite{Brandenburg:2005he}.%
%
%\footnote{The differences between the two scenarios shown in
%  Fig.~\ref{Fig:Fingerprint} become smaller for larger values of
%  $m_{\Bi} / m_{\st}$.}
%
\footnote{There is a caveat: If $\mgr < m_{\ax} < m_{\st}$ and
  $\Gamma(\stau \to \axino\,X) \gg \Gamma(\stau \to \gravitino\,X)$,
  one would still find the distribution shown in the upper panel of
  Fig.~\ref{Fig:Fingerprint}. The axino would then eventually decay
  into the gravitino LSP and the axion.}

To be specific, with $10^4$ analyzed stau NLSP decays, we expect about
165$\pm$13 (stat.) events for the $\axino$ LSP and about 100$\pm$10
(stat.) events for the $\gravitino$ LSP~\cite{Brandenburg:2005he},
which will be distributed over the corresponding
($x_\gamma$,\,$\cos\theta$)-planes shown in
Fig.~\ref{Fig:Fingerprint}. In particular, in the region of
$x_{\gamma}\gtrsim 0.8$ and $\cos\theta \lesssim -0.3$, we expect
about 28\% of the 165$\pm$13 (stat.) events in the $\axino$ LSP case
and about 1\% of the 100$\pm$10 (stat.) events in the $\gravitino$ LSP
case.  These numbers illustrate that $\Order(10^4)$ of analyzed stau
NLSP decays could be sufficient for the distinction based on the
differential distributions. To establish the feasibility of this
distinction, dedicated studies including details of the detectors and
the additional massive material will be
crucial~\cite{Hamaguchi:2006vu}.

% __________________________________________________________________
\subsection{Probing the Peccei-Quinn Scale \boldmath$f_a$ and \boldmath$m_{\axino}$}
\label{sec:AxinoPQScale}
% __________________________________________________________________

If $\axino$ is the LSP and $\stau$ the NLSP, 
the analysis of the 2-body decay $\stau\to\tau\axino$ will allow us to
probe the Peccei-Quinn scale $f_a$ and the axino mass $m_{\axino}$.
In fact, the measurement of $\taustau$~(\ref{Eq:Axino2Body}) with
methods described in Sect.~\ref{sec:GravitinoExperiments} leads to the
following estimate of the Peccei-Quinn scale
$f_a$~\cite{Brandenburg:2005he}:
\bea
  f_a^2 
  &\simeq&
  {\xi^2\,C_{\rm aYY}^2} 
  \left(10^{11}\,\GeV\right)^2
  \Big(1-\frac{{m_{\axino}^2}}{{m_{\stau}^2}}\Big)
  \left(\frac{\tau_{\stau}}{25~\mathrm{s}}\right)\, 
\nonumber\\
  && \times
  \left(\frac{m_{\stau}}{100\,\GeV}\right)
  \left(\frac{m_{\Bino}}{100\,\GeV}\right)^2
  \ ,
\label{Eq:PQ_Scale}
\eea
which can be confronted with $f_a$ limits from astrophysical axion
studies and axion searches in the
laboratory~\cite{Yao:2006px,Sikivie:2006ni,Raffelt:2006rj,Raffelt:2006cw,Ringwald:2003nsa}.
Indeed, we expect that $m_{\st}$ and $m_{\Bino}$ will already be known
from other processes when the $\stau$ NLSP decays are analyzed;
cf.~Sect.~\ref{sec:GravitinoExperiments}. The dependence on $m_{\ax}$
is negligible for $m_{\ax}/m_{\st}\lesssim 0.1$.  For larger values of
$m_{\ax}$, the $\stau$ NLSP decays can be used to determine $m_{\ax}$
from the kinematics of the 2-body decay, i.e., from a measurement of
the energy of the emitted tau $E_\tau$,
\begin{align}
  m_{\axino} =
  \sqrt{{m_{\st}^2}+{m_\tau^2}-2{m_{\st} E_\tau}} 
  \ ,
\label{Eq:Axino_Mass} 
\end{align}
with an error governed by the experimental uncertainties on $m_{\st}$
and $E_\tau$. As is evident from~(\ref{Eq:AxinoDensityTP})
and~(\ref{Eq:AxinoDensityNTP}), the determination of both the
Peccei--Quinn scale $f_a$ and the axino mass $m_{\ax}$ is crucial for
insights into the cosmological relevance of the axino LSP.

% __________________________________________________________________
\section{Conclusion}
\label{sec:Conclusion}
% __________________________________________________________________
%
\begin{table*}
  \caption{Supersymmetric dark matter candidates, their identity, and key properties. With the listed production mechanisms, $\Omega_{\LSP}=\Omega_{\CDM}$ is possible. The respective production leads typically to a cold, warm, or hot dark matter component as indicated. Quantities marked with `(?)' seem to be unaccessible in light of the current understanding of cosmological constraints within a standard thermal history.}
\label{tab:SUSYDMCandidates}
\begin{tabular}{lllllll}
\hline\noalign{\smallskip}
LSP & 
identity &
mass &
interactions &
production &
constraints &
experiments\\
\noalign{\smallskip}\hline\noalign{\smallskip}
$\widetilde{\chi}^0_1$ & 
lightest neutralino &
$\Order(100\,\GeV)$ &
g, g', $y_i$ &
therm.~relic &
$\leftarrow$ cold &
indirect searches
\vspace{0.1cm}
\\
&
(spin 1/2)&
& 
weak &
&
&
direct searches
\vspace{0.1cm}
\\
&
mixture of&
&
{\tiny $M_{\mathrm{W}}\sim 100~\GeV$} &
&
&
collider searches
\vspace{0.1cm}
\\
&
$\Bino$, $\Wino$, $\HiggsinoUp$, $\HiggsinoDown$&
&
&
&
&
%\vspace{0.1cm}
\\
\noalign{\smallskip}\hline\noalign{\smallskip}
$\widetilde{G}$ &
gravitino &
eV--TeV &
$(p/\MPl)^{n}$ &
therm.~prod. &
$\leftarrow$ cold &
$\stau$ prod. at colliders
\vspace{0.1cm}
\\
&
(spin 3/2)&
&
extremely weak &
NLSP decay &
$\leftarrow$ warm &
+ $\stau$ collection 
\vspace{0.1cm}
\\
&
superpartner&
&
{\tiny $\MPl = 2.4\!\times\! 10^{18}\,\GeV$} &
&
&
+ $\stau$ decay analysis
\vspace{0.1cm}
\\
&
of the graviton&
&
&
&
BBN&
$\hookrightarrow$ $\mgravitino$, $\MPl$ (?)
%\vspace{0.1cm}
\\
\noalign{\smallskip}\hline\noalign{\smallskip}
$\widetilde{a}$ &
axino &
eV--GeV &
$(p/f_a)^{n}$ &
therm.~relic. &
$\leftarrow$ hot/warm &
$\stau$ prod. at colliders
\vspace{0.1cm}
\\
&
(spin 1/2)&
&
extremely weak &
therm.~prod. &
$\leftarrow$ cold/warm &
+ $\stau$ collection
\vspace{0.1cm}
\\
&
superpartner&
&
{\tiny $f_a\gtrsim 10^{9}\,\GeV$} &
NLSP decay &
$\leftarrow$ warm/hot&
+ $\stau$ decay analysis 
\vspace{0.1cm}
\\
&
of the axion&
&
&
&
BBN&
$\hookrightarrow$ $m_{\ax}$ (?), $f_a$
%\vspace{0.1cm}
\\
\noalign{\smallskip}\hline
\end{tabular}
\end{table*}

Dark matter is strong evidence for physics beyond the Standard Model.
Extending the Standard Model with SUSY, an electrically neutral and
color neutral LSP becomes a dark matter candidate for conserved
R-parity. I have shown that the neutralino $\neutralino$, the
gravitino $\gravitino$, and the axino $\axino$ can be the LSP and as
such explain the non-baryonic dark matter in our Universe.  The
neutralino $\neutralino$ is already part of the MSSM which provides a
solution of the hierarchy problem and allows for gauge coupling
unification. Being the superpartner of the graviton and the gauge
field associated with supergravity, the gravitino $\gravitino$ is
equally well motivated with a mass $\mgr$ that reflects the SUSY
breaking scale. As the superpartner of the axion, also the axino
$\axino$ appears naturally once the strong CP problem is solved with
the Peccei--Quinn mechanism in a SUSY setting.

While mass values and interactions can be very different for the
$\neutralino$, $\gravitino$, and $\axino$, I have illustrated for each
of these LSP candidates that natural regions in the parameter space
exist in which $\Omega_{\LSP}=\Omega_{\CDM}$. These regions are
limited by bounds from electroweak precision observables, B-physics
observables, Higgs and sparticle searches at LEP, and by BBN
constraints. The constraints from $\Omega_{\CDM}$ and BBN also imply
serious upper limits on the reheating temperature after inflation
$\TR$ which can be relevant for models of inflation and baryogenesis.

Most promising are the experimental prospects in the case of the
$\neutralino$ LSP. Being a WIMP, the $\neutralino$ LSP should be
accessible in direct and indirect dark matter searches. Indeed, first
hints might have already been found in the EGRET
data~\cite{deBoer:2005bd,Elsaesser:2004ap}. With ongoing indirect
searches, the increasing sensitivity of direct searches, and the
advent of the LHC at which $\neutralino$ dark matter could be
produced, we will be able to test whether these hints are indeed the
first evidence for the existence of SUSY dark matter. While an excess
in missing transverse energy is expected to be the first evidence for
SUSY at the LHC already within the next three years, the
identification of the $\neutralino$ being the LSP will require the
reconstruction of the SUSY model realized in nature. If superparticles
are within the kinematical reach, precision studies at the ILC will be
crucial for this endeavor.

In the $\gravitino/\axino$ LSP scenarios with conserved R-parity, no
dark matter signal should appear in direct or indirect searches.
However, since an electrically charged LOSP such as the $\stau$ is
viable in the $\gravitino/\axino$ LSP scenarios, (quasi-) stable
$\stau$'s might occur as muon-like particles instead of an excess in
missing transverse energy. Indeed, an excess of (quasi-) stable
$\stau$'s could appear as an alternative first evidence for SUSY at
the LHC in the next three years. Because of the severe upper limits on
the abundance of stable charged particles~\cite{Yao:2006px}, one would
then expect that the $\stau$ is the NLSP that decays eventually into
the $\gravitino/\axino$ LSP or that R-parity is broken. A distinction
between these scenarios will require the analysis of the $\stau$
decays. For this challenge, the ILC with its tunable beam energy seems
crucial~\cite{Hamaguchi:2004df,Feng:2004yi,Martyn:2006as,Hamaguchi:2006vu,Martyn:2007mj}.

Table~\ref{tab:SUSYDMCandidates} presents an overview of the SUSY dark
matter candidates discussed in this review. As the LSP, 
each of them---the lightest neutralino $\neutralino$, the gravitino
$\gravitino$, or the axino $\axino$---could provide $\Omega_{\CDM}$
and could be produced and identified at colliders in the near future.

% __________________________________________________________________
\bigskip
% __________________________________________________________________

I would like to thank the organizers of SUSY 2007 for inviting me to
an exciting and stimulating conference. I am grateful to
A.~Brandenburg, L.~Covi, A.~Freitas, K.~Hamaguchi, G.~Panotopoulos,
T.~Plehn, J.~Pradler, L.~Rosz\-kowski, S.~Schilling, N.~Tajuddin,
Y.~Y.~Y.~Wong, D.~Wyler, and M.~Zagermann for valuable discussions and
collaborations on the topics covered in this review.

% __________________________________________________________________
% 
%\bibliographystyle{epj}
%\bibliography{biblio}
%

%
% __________________________________________________________________

\end{document}